\documentclass[ACS,STIX2COL]{WileyNJD-v2}

\articletype{Full Article}%

\received{ }
\revised{ }
\accepted{ }

\usepackage{graphicx}
\usepackage{amsmath,amsfonts}
\usepackage{tabu}
\usepackage{pdflscape}
\usepackage{makecell}
\usepackage{booktabs}
\usepackage{pdfpages}
\usepackage{epstopdf}
\usepackage{multirow}
\usepackage{enumitem}
\usepackage{wrapfig}
\usepackage{textcomp}
\usepackage{tablefootnote}
\usepackage{blindtext}
\usepackage{subfig}
\DeclareGraphicsExtensions{.jpg,.png,.eps}

\begin{document}

\title{Rapid, remote and low-cost finger vasculature mapping for heart rate monitoring}

\author[1]{Akhil Kallepalli*}
\author[2]{David B. James}
\author[2]{Mark A. Richardson}

\authormark{Kallepalli \textsc{et al}}

\address[1]{\orgdiv{School of Physics and Astronomy}, \orgname{University of Glasgow}, \orgaddress{\state{Glasgow}, \country{United Kingdom}}}

\address[2]{\orgdiv{Centre for Electronic Warfare, Information and Cyber}, \orgname{Cranfield University}, \orgaddress{\state{Defence Academy of the UK, Shrivenham}, \country{United Kingdom}}}

\corres{* Direct all correspondence to AK \\ \email{Akhil.Kallepalli@glasgow.ac.uk}}

\presentaddress{* Optics Group, School of Physics and Astronomy, Level 3, Advanced Research Centre, University of Glasgow, Glasgow G12 8QQ}

\abstract{Today's diagnostics include devices such as pulse oximeters, blood pressure monitors, and temperature measurements. These devices provide vital information to medical personnel when making treatment decisions. Drawing inspiration from the fundamental utility of pulse oximeters, we present a methodology for a robust low-cost approach to imaging subsurface vasculature and monitoring heart rate. The approach uses off-the-shelf equipment, set up in free space without physical contact and exploits the nature of the interaction between light at near-infrared wavelengths with tissue. Image processing algorithms extract heart rate information from the snapshot and video sequence captured at a stand-off distance. The method can be applied in a room with ambient light and remains robust to scenarios comparable to medical situations. This research sets the platform for future diagnostic devices based on imaging systems and algorithms for non-contact point-of-care investigations.}



\maketitle

Any medical scenario requires a quantitative assessment of an anatomical metric. For diagnostics, vital signs such as heart rate, temperature, oxygen saturation inform medical decisions from outpatient visits to monitoring patient health during life-saving surgeries. Further, some of these signs have become a key indicator of health during the ongoing COVID-19 pandemic \cite{Hussain2021}. In the latter context, non-contact measurement of temperature became a key aspect and utility of such devices was, and remains, undeniable. Non-contact methods offer a multitude of advantages such as ease of access, minimal training, reduction of equipment surrounding the patient, sterilisation of equipment prior to reuse, elimination of qualitative determinations, improved comfort to the patient, and many others \cite{Campbell2018}. 

Motivated by the need and advantages of remote detection of vital signs, we explore a simple, easy-to-use system operating in a free space arrangement for detection and monitoring vascular activity. The research provides an approach that leverages a laser-based, non-contact optical system to monitor vascular activity. The interaction of tissue with light at near-infrared wavelength is well researched. In the biophotonics domain, it is well understood that optical diagnostics exploit the visible and near-infrared wavelengths \cite{BoudouxFundamentals2017_1}. This window is limited by ultraviolet wavelengths ($<400$ nm due to the photo-chemical reaction that causes melanogenesis) and the near-infrared on the other end ($>1020$ nm due to water absorption) \cite{Jacques2013}. This non-ionising, skin-safe range of the spectrum offers innumerable diagnostic capabilities and quantitative imaging methods. 

Years of fundamental and applied research of photonics and optics show the utility of optical methods for biophotonics \cite{Goda2019}. The applications of optical technologies in medical research are innumerable, could fill books and the state-of-the-art capabilities are constantly being redefined \cite{Li2022,Greisen2022,Sowers2022}. While these technologies are opening avenues of fundamental and life-saving tools, there remains a constant search of low-cost methods that can make diagnostics accessible the world over; an objective of distinct priority of the research presented in this article. 

\section{Background}
\label{sec:Background}
In this research, the objective is to image the subsurface vasculature and monitor its behaviour. While Monte Carlo (MC) assessment of light transport through the skin layers and into the blood vessel was assessed in early research \cite{Nilsson1998}, Boles and Chu (1997) \cite{Boles1997} illustrated the possibility of personal authentication using images of the human palm. This initiated research in both the biomedical \cite{Fujimasa2000,Cuper2013,Gao2016,Jacques2013} and biometric \cite{Wang2006,Hashimoto2006,Mulyono2008,Lee2011,Yang2012,Qin2013,Kang2014,Yang2014,Damavandinejadmonfared2015,Gupta2015,Tien2015,Kono2015,Pham2015,Nguyen2017,Lee2017,Matsuda2017} domains.
The current research draws inspiration from biometric identification methods and extends it to identification and monitoring of blood vessels for non-contact heart rate measurement. 

The applications of imaging blood vessel networks has seen interesting applications such as banking and security; vessel networks are unique to each individual. Subsequently, convolutional neural networks (CNN) also found application in finger vein recognition to avoid spoofing using carbon ink images that can be used to defeat a vein recognition system \cite{Nguyen2017}. A combination of ‘liveness detection’ to this system was done, without CNN, by using imaging longer than a single snapshot as an anti-spoofing strategy \cite{Lee2017}. 

Subsequent work focused on infrared imaging at different body sites. These methods suffered from interference from ambient conditions and human body conditions when using the far-infrared wavelengths, while the near-infrared transmission was inconsistent due to skin and hair attenuation \cite{Wang2006}. The transmissive arrangement was used extensively \cite{Mulyono2008,Lee2011,Yang2012,Gupta2015,Pham2015} while other methods attempted to use data fusion combining visible and near-infrared transmission images for imaging the veins and the dorsal texture of the skin for identification \cite{Yang2014} or combining reflectance and transmission near-infrared data \cite{Tien2015}. A key review article in this domain was presented by Hashimoto (2006) mentioning the advantages of non-invasive imaging, safety from forgery or theft due to the veins being hidden under the surface and the stability of vasculature allowing simplistic methods to capture these patterns \cite{Hashimoto2006}. The article highlights the advantages of transmissive imaging and using vein profiles in greyscale images as a feasible option. These vein profiles are the mainstay of the current study.

Vasculature mapping is important in the biomedical domain, as much as it is used in the biometrics domain. One of the many instances, where it is a good tool, is the usage of near-infrared light to visualise the subsurface vasculature for simpler withdrawal of blood in children \cite{Cuper2013}. The developed device, VascuLuminator, also considers the implications of skin types and performs well under ambient conditions, claiming an overall drop of 13-20\% of first attempt failures to extract blood from children under the age of six \cite{Cuper2013}. 

The objective encapsulates a simple inquiry -- using lasers and a low-cost imaging system to monitor the cardiac activity of a target in motion. Of the many algorithms, a choice was made to apply two methods: maximum curvature from image profiles \cite{Miura2005} and repeated line tracking \cite{Miura2004}. The choice was supported by the high degrees of accuracy when compared to other algorithms used for similar biometric analyses \cite{Yang2012,Qin2013}. 

\section{Optical Investigation}
\label{sec:Optic_SkinTransm}
The experimental conditions were constant for all the participants \footnote{The ethics approval for conducting this research was obtained prior to experimental procedures from the University's ethics committee (CURES/2208/2017). Written consent from the participants was collected prior to experimentation.}. The participants remained comfortably seated to acclimatise to the room for 15 minutes, prior to data collection. The image sequence data collected from the participants, wherein the participant holds their finger steady and a sequence of frames where the finger moves, perpendicular to the beam's direction. The intention with this data acquisition is to test the algorithm for detection and monitoring to remain robust when the finger moves inherently and differently to each individual. During data capture, an off-the-shelf pulse oximeter (Berry Pulse Oximeter, Type: BM2000A) was attached to the participant's index finger on the other hand for measuring their heart rate. 

\begin{figure} [ht] 
\begin{center}
\includegraphics[width=0.85\columnwidth]{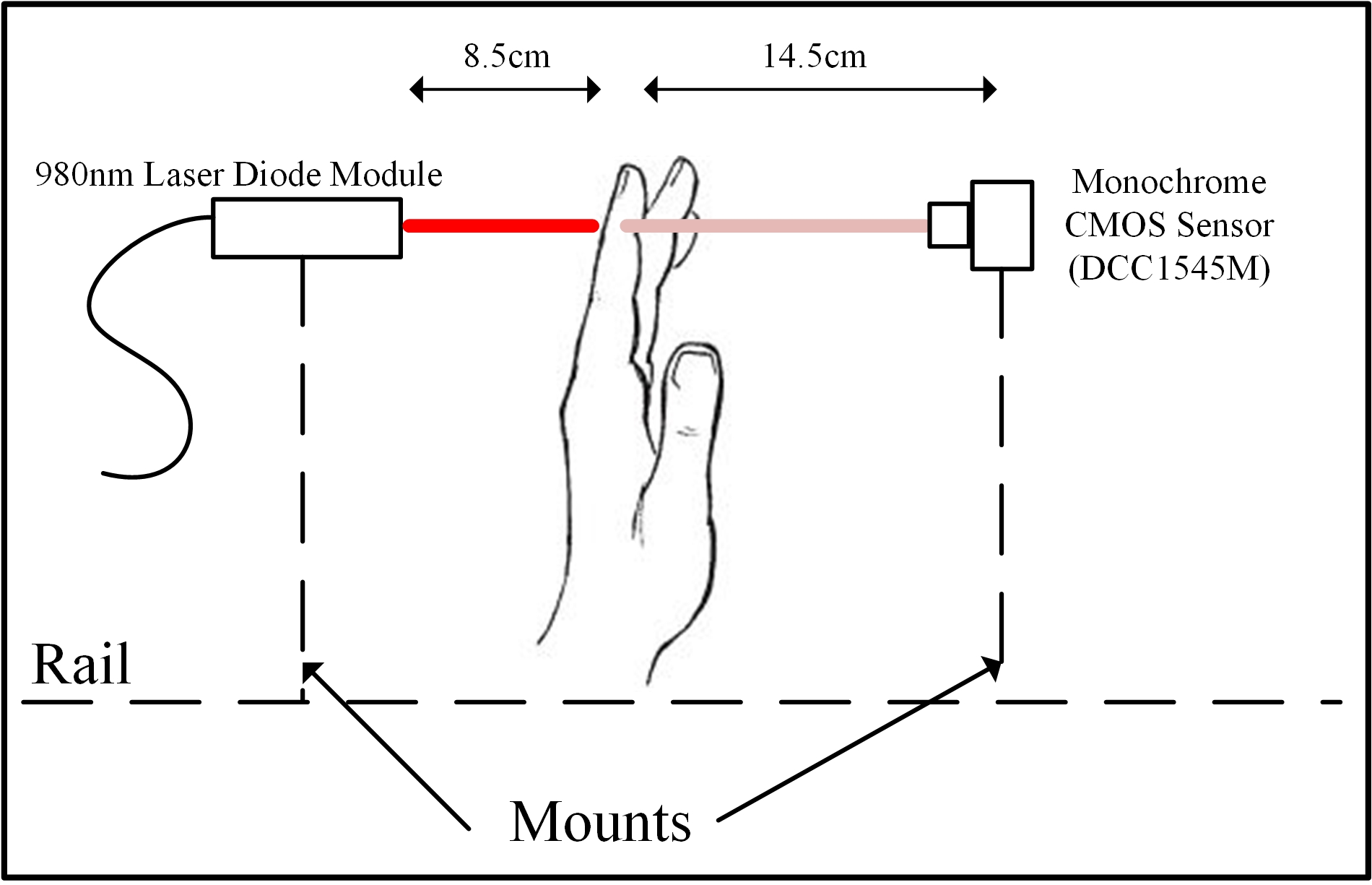}
\end{center}
\caption{The imaging system comprises of a monochrome camera (DCC1545M, Thorlabs Inc.) and a machine vision lens (MVL4WA, Thorlabs Inc.). The sources used are continuous wave near-infrared (CPS980, Thorlabs, Inc.) and red (CPS670F, Thorlabs, Inc.) laser diode modules. The simple setup exploits the interaction of light with human tissue to measure the heart rate.}
\label{fig:Experimental_Design}
\end{figure}

The study investigates the \textit{in vivo} transmission of red and near-infrared light through the index finger of 12 participants, spanning across Fitzpatrick's dermatological classification of skin types \cite{Fitzpatrick1988}. To understand the implications of skin tone, if any, in these experiments, the participants were placed in three groups along Fitzpatrick's scale; 4 participants in group 1 (I and II on Fitzpatrick's classification scale), 3 participants in group 2 (III and IV on Fitzpatrick's classification scale) and 5 participants in group 3 (V and VI on Fitzpatrick's classification scale). However, this grouping is not factored into the experiment in any way except for a qualitative comparison after data processing and analysis. Due care must be taken when drawing inferences from this study as an $n=12$ size cannot be used for larger generalisations. 

The focus of this research is the detection of the vascular bundle and the determination of heart rate. We chose off-the-shelf equipment and a relatively simple approach for data collection, followed by data analysis that can be integrated into the data collection approach for real-time monitoring in the future. The experiments were done using an imaging system comprised of a monochrome camera and lens system, and skin-safe lasers.The intensity of transmitted light is imaged while using a single wavelength laser source, removing the directional uncertainty that occurs when using diffuse light sources \cite{Miura2004,Miura2005}. The light sources used in this study are continuous wave near-infrared (CPS980, Thorlabs, Inc.) and red (CPS670F, Thorlabs, Inc.) laser diode modules. The laser powers (Class 3R) were confirmed with the Lasermet ADM 1000 power meter to ensure the incidence of the skin is substantially lower than the maximum permissible exposure (MPE) limit. The light is incident on the dorsal side of the finger and the transmitted light is imaged directly behind the finger by a USB-operated monochrome CMOS camera (DCC1545M, Thorlabs, Inc.). The camera is paired with a machine vision lens (MVL4WA, Thorlabs, Inc.), of focal length 3.5 mm operating with an \textit{f}-number of 1.4 (\textit{f}/1.4), to capture the transmittance in a light-tight room (Figure \ref{fig:Experimental_Design}). 

The laser and the camera is stationary through the course of the experiment, with the only movement being the participants' horizontal hand movement. The data collected includes horizontal movement and any minor movements inherent to each participant. Combined, all the motion in the imaged region are the challenges that this study will overcome to image the subsurface vasculature and monitor the heart rate. 

\section{Vein Mapping and Monitoring}
\label{sec:Vein}
\begin{figure}[h]
\begin{center}
\includegraphics[width=0.9\columnwidth]{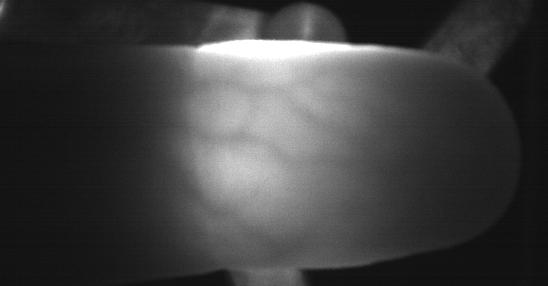}
\end{center}
\caption{Near-infrared wavelengths are absorbed by the blood vessels while the remaining light scatters through and propagates to the imaging system behind the finger. The resulting ``shadow'' image allows mapping of vasculature and, in this study, monitoring them as a function of expanding and contracting blood vessels.}
\label{fig:Image_Vein}
\end{figure}

\begin{figure*} [t]
\centering
\includegraphics[width=1.6\columnwidth]{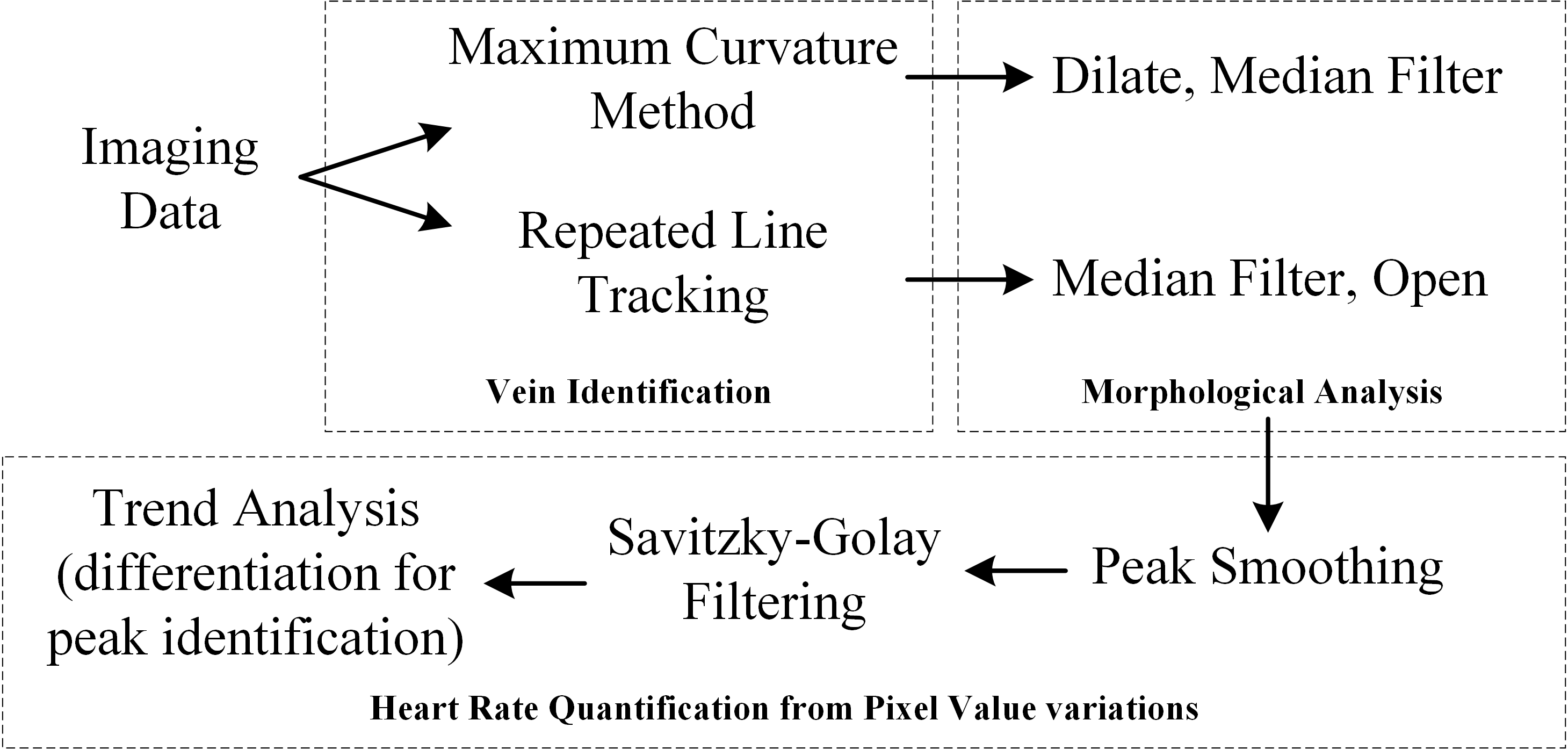}
\caption{The methodology is illustrated here; The data collected from the imaging system is processed using two algorithms for vein mapping: maximum curvature and repeated line tracking methods. The results of these methods are processed with morphological analysis (using dilation and open filters) with noise removed using median filters. The results of this step are assessed for heart rate measurement using peak smoothing, Savitzky-Golay filtering and trend analysis.}
\label{fig:Algorithm_Workflow}
\end{figure*}

The differential interaction of light with tissue and blood vessels at different wavelengths results in `shadow' image (Figure \ref{fig:Image_Vein}) wherein the vessel network can be mapped and quantified. The image processing and analysis was done using MATLAB (R2018b). The images are processed using two algorithms (Figure \ref{fig:Algorithm_Workflow}) for performance analysis and comparison -- repeated line tracking and maximum curvature methods. 

\begin{figure}[t]
\centering
\includegraphics[width=0.9\columnwidth]{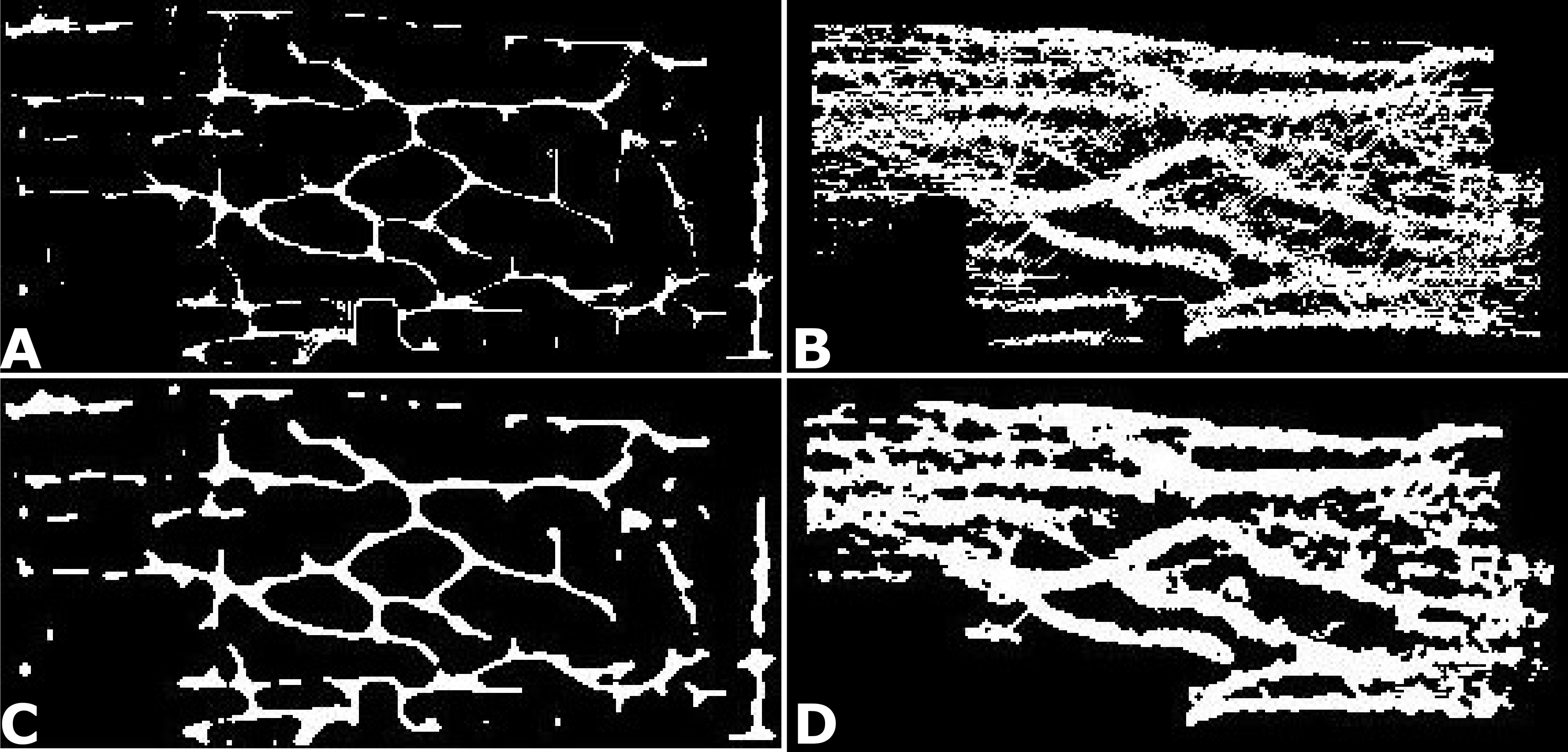}%
\caption{The image processing in this workflow use the intrinsic properties of the vein identification algorithm outcome to improve the vascular maps. These include using morphological approaches in pixel neighbourhoods and treating random identifications in the vessel detection algorithms as noise. Consequently, the results of the maximum curvature method (A) are dilated and median filter is used to remove noise (C). For the repeated line tracking method (B), a median filter is used to remove noise and morphologically eroding and dilating the pixels to enhance the maps (D).}
\label{fig:Step2_ImageProcessing}
\end{figure}

The maximum curvature method uses approaches to localise the finger boundary from the background \cite{Lee2009}. In the maximum curvature method, the variation of pixel values across blood vessels are used. As the blood vessels absorb near-infrared light, they create darker profiles and therefore, a lower value in the vertical profile of pixel values (in a column of pixels in the image). The subsequently created `valley' of pixel values not only represents the blood vessel size and location, but also also oscillates as a function of the heartbeat (ref. to Miura \textit{et al.} 2005 \cite{Miura2005} for a detailed description of the algorithm). This approach, iterated over an entire image, results in a vascular bundle map in each image. Once the finger region is localised and the features are extracted, morphological analyses are applied. Morphological analyses comprise a broad range of image processing operations that use the information regarding the shapes in the image and adjust every pixel based on its neighbourhood. In this case, we use a dilation operation to the results of the maximum curvature method as a step toward extending the broken vessel mapping. Subsequently, the misidentified pixels are removed with a median filter (Figure \ref{fig:Step2_ImageProcessing}). The resulting outputs of morphological analyses is illustrated in Figure \ref{fig:Step2_ImageProcessing} (A, C). 

On the other hand, the repeated line tracking approach is used to detect features in a relatively random manner. The algorithm randomly `picks' pixels and explores the neighbourhood for similar values. When found, the algorithm extends the similar pixels and joins them to create lines (vasculature, in this case). Due to the random nature of this algorithm, it is far more susceptible to noise. After the vessels are identified and extracted, a median filter is used to remove the misidentified pixels. These, with a specific filter window, removes pixels that are not a part of an extended vessel from the repeated line tracking method. As a result, fewer noisy pixels and an extended vasculature map can be visualised (Figure \ref{fig:Step2_ImageProcessing} (B, D)).

\begin{figure*} [htp]
\begin{center}
\includegraphics[width=0.8\columnwidth]{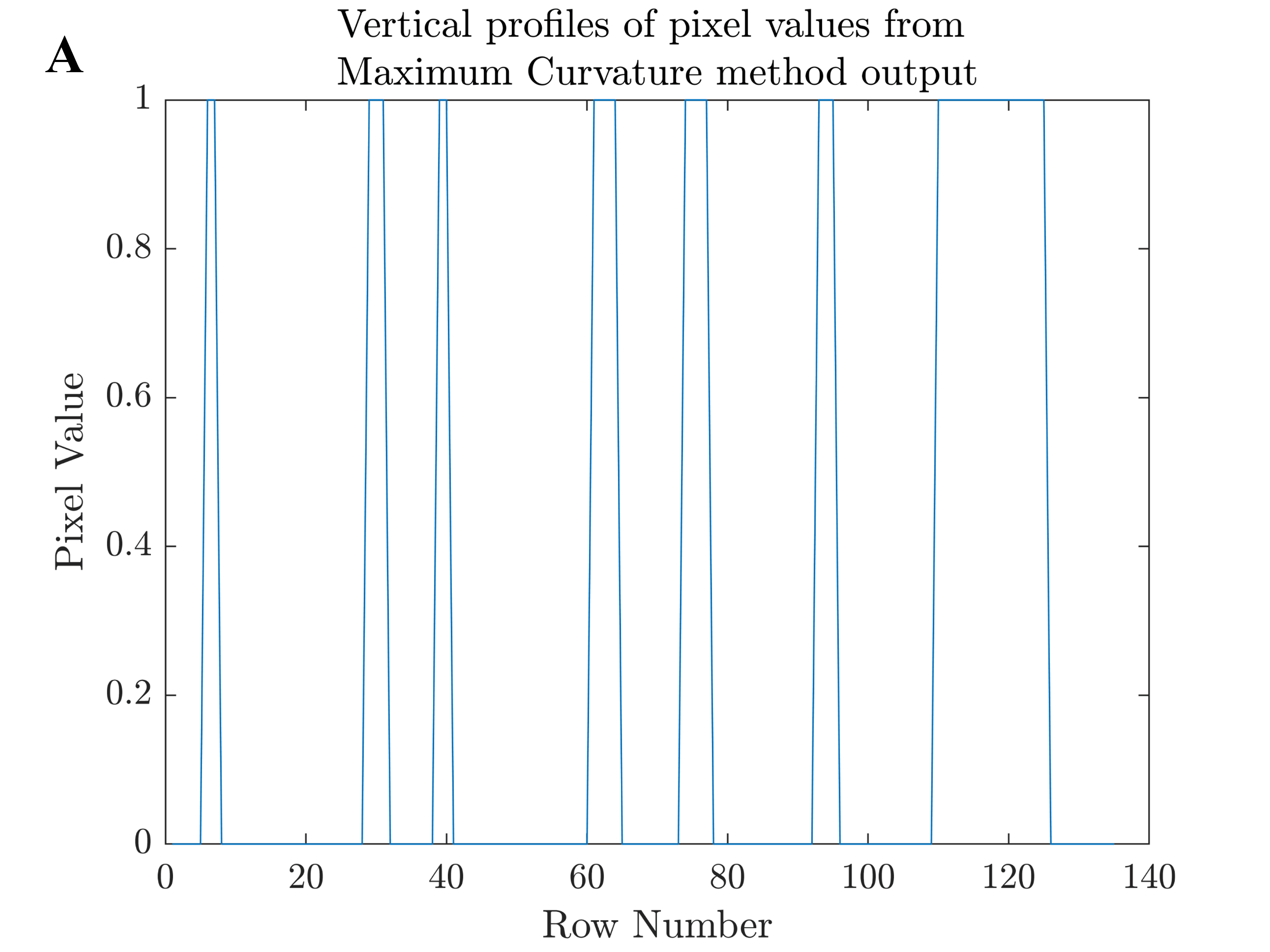}
\includegraphics[width=0.8\columnwidth]{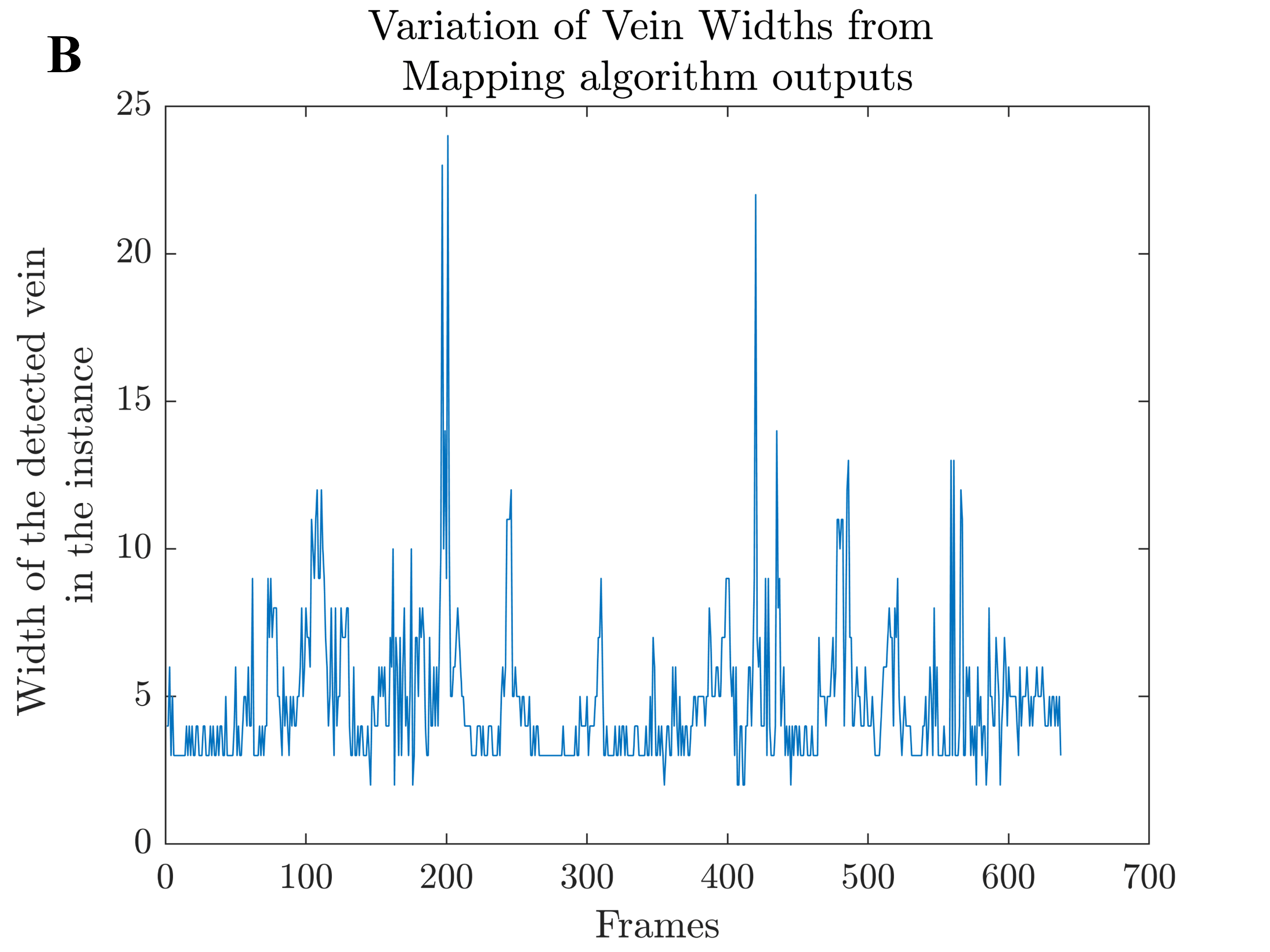} \\
\includegraphics[width=0.8\columnwidth]{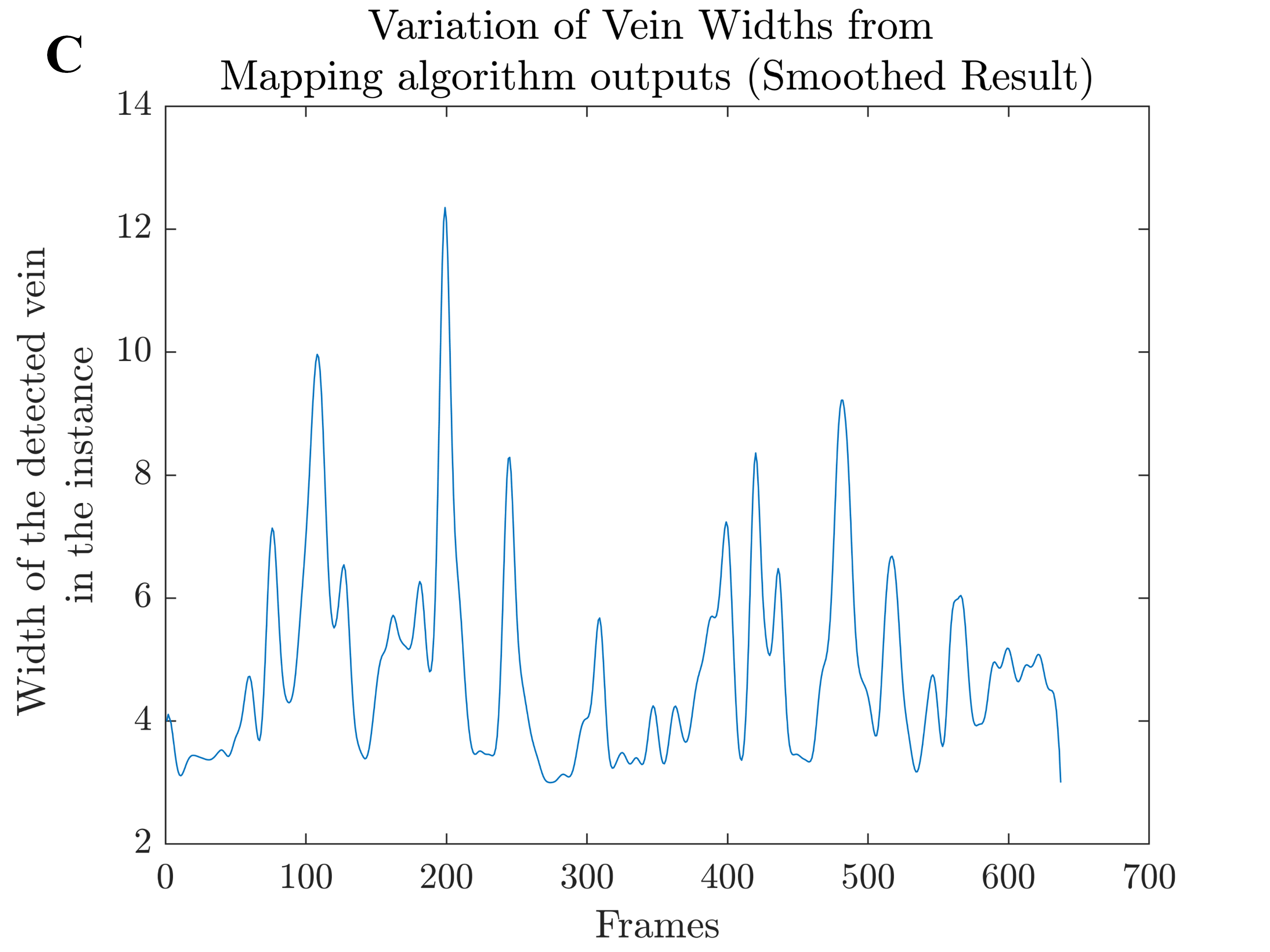}
\includegraphics[width=0.8\columnwidth]{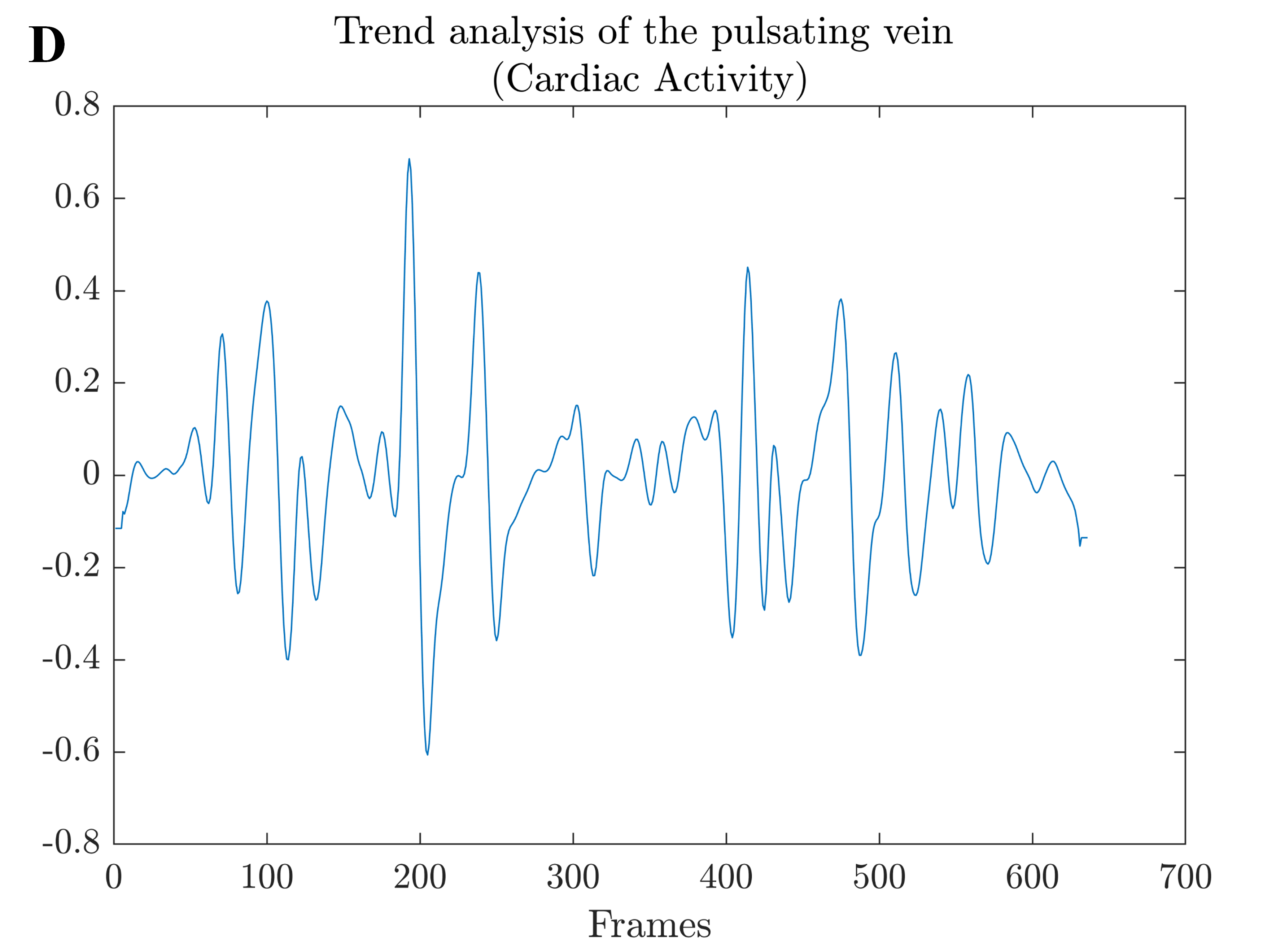} \\
\includegraphics[width=1.6\columnwidth]{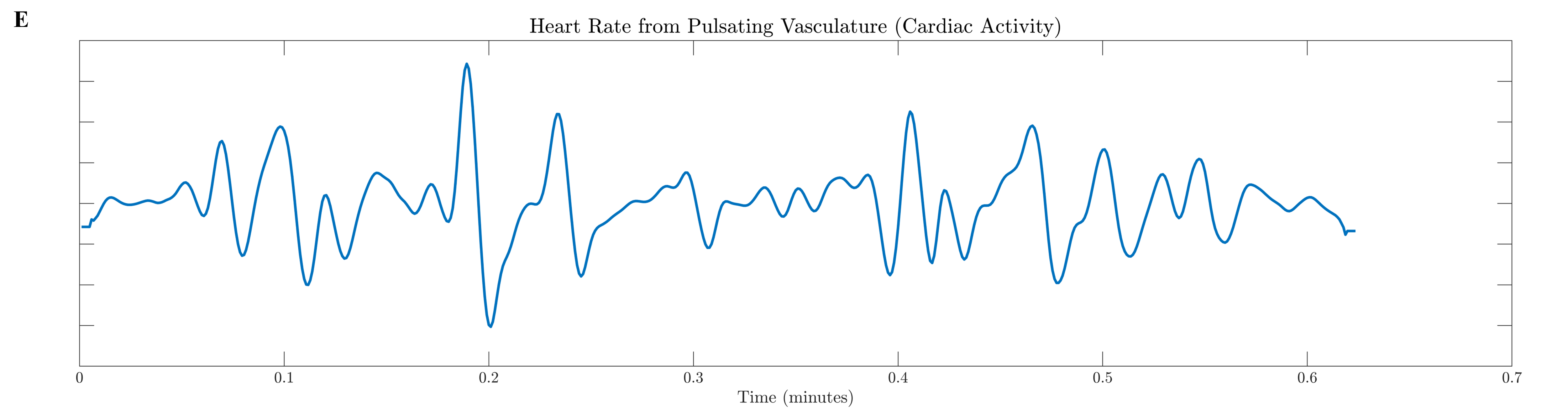}
\end{center}
\caption{Once the image data is captured, vascular maps are generated and improved with morphological analyses, the data analysis is done using filtering and trend analysis. The positions of the veins are detected using vertical profiles across binary images (A). The choice of veins for monitoring the heart rate is made approximately from the central part of the profile as this has the highest probability for an accurate identification. Once identified, the vein thickness is measured in terms of number of pixels across the entire acquisition (B). This output is smoothed to give (C), which is further smoothed using Savitzky-Golay filtering. Subsequent trend analysis results in the algorithm output, as shown in (D). The frames are converted to equivalent time periods to illustrate the cardiac activity (E) and successfully compared with pulse oximeter readings. Note that the result is not fitted to a sinusoidal pattern, which is done by commercial devices.}
\label{fig:Step3_ImageProcessing}
\end{figure*}

At this stage of processing, the images have been converted into binary images consisting of vasculature maps, along with information of vertical profiles from which the maps are created. Once again, vertical profiles are drawn across the image. From these profiles, the locations of the blood vessels in each frame of the video sequence are identified based on the location of the peaks (Fig. \ref{fig:Step3_ImageProcessing} (A)). The methodology picks the location of the blood vessel from each frame of the video sequence using the information from these vertical profiles and plots the degree of variation of vasculature width in all the frames of the dataset (Fig. \ref{fig:Step3_ImageProcessing} (B)). This step is the mainstay for removing motion artefacts from the monitoring process as every frame of the video is treated as an individual image. For every image, blood vessels are identified, morphological analyses are performed to enhance the vasculature map and the trend of the blood vessel in terms of pixel values is quantified. 

The variation trend is smoothed with a Savitzky-Golay filter (Figure \ref{fig:Step3_ImageProcessing} (C)). The Savitzky–Golay filter provides the advantage of preserving the position and width of the peaks by fitting subs-sets of adjacent data points with minimal distortion of the trend. This is important as small variations between frames can be smoothed but the overall trend must be preserved. Finally, the trend function is differentiated to obtain a `rate of change' of the blood vessel's width, resulting in a measure of the heart rate (Figure \ref{fig:Step3_ImageProcessing} (D)). Since the frame rate of the camera acquisition is known, the heart rate can now be represented as a function of time (Figure \ref{fig:Step3_ImageProcessing} (E)), with the peaks representing the heartbeat. The number of peaks in the final output of the algorithm workflow represents the heart rate and, confirmed by a pulse oximeter, they are accurate within $+/-$ 5 beats per minute (bpm). 

The combination of the smoothing algorithm and observing the rate of change of the vessel thickness is key in removing the errors that could arise from the motion of the finger. It is important to note that the outcome of the workflow is not post-processed or fit to a sinusoidal function as is done by commercial devices. Although this is a simple step, the results are kept in this format to include the vasculature size variations that are inherently a part of this heart rate measure. 

\begin{figure*} [htp]
\begin{center}
\includegraphics[width=1.8\columnwidth]{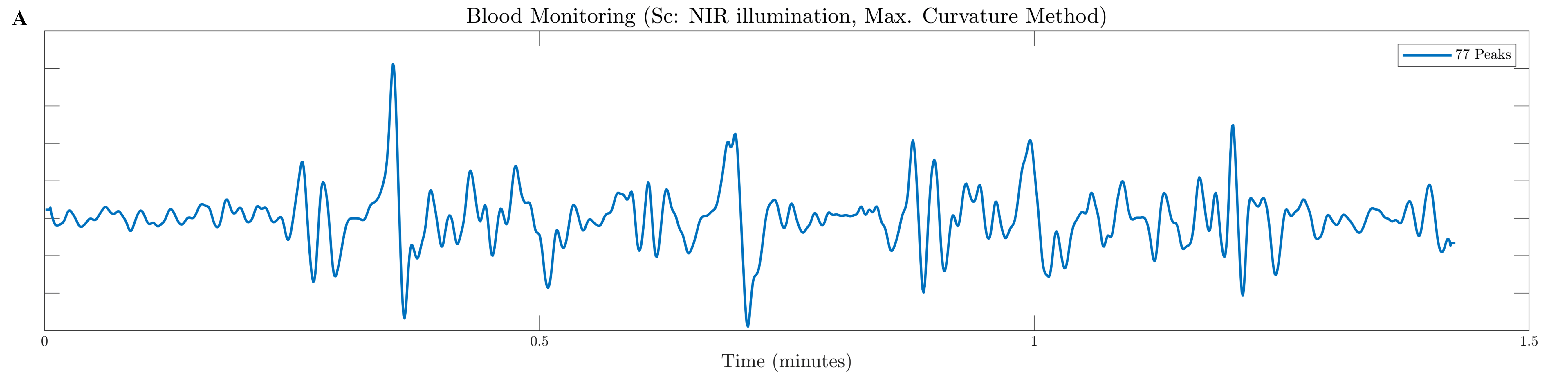} \\
\includegraphics[width=1.8\columnwidth]{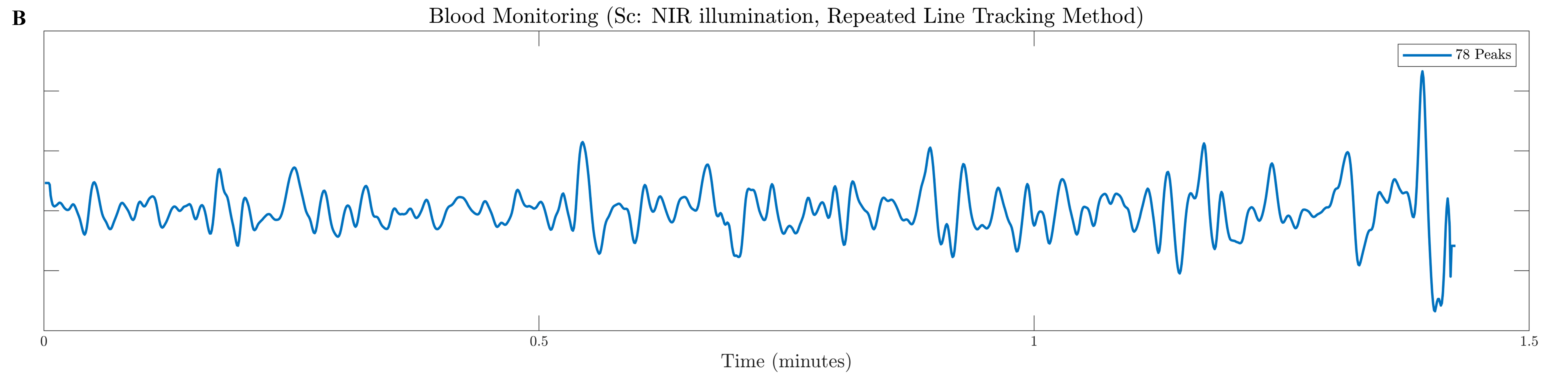} 
\end{center}
\caption{The vasculature monitoring results of the complete methodology for USP001 compares the results from both pattern mapping algorithms. The results show a clear trend but a better consistency of sinusoidal fit using the repeated line tracking method (B). Both methods detect 77 peaks, i.e. beats within the duration of capture of less than 90 seconds. The resting heart rate was confirmed using a pulse oximeter with a margin of $+/-$ 5 beats per minute.}
\label{fig:Results_1}
\end{figure*}

\section{Results and Discussions}
\label{sec:Disc}
The method described above focuses on imaging finger vasculature and extracting their pattern. Subsequently, the patterns and the vasculature would be used to monitor cardiac activity. The left index finger was used for imaging. An image of the near-infrared illuminated finger, the output of the maximum curvature and repeated line tracking algorithms is given in the Supplementary Material through figures S1 -- S12. 

When comparing the results (from before the image processing step) qualitatively, the maximum curvature method by far outperforms the repeated line tracking method, qualitatively. The binary images generally have fewer erroneous identifications when using the maximum curvature method. The repeated line tracking method is very susceptible to noise and this is evident (Fig. \ref{fig:Step2_ImageProcessing} and Supplementary figures \textbf{S1 -- S12}). However, the contrast with which the veins are identified is far more significant in the repeated line tracking method. The results in the figures S1--S12 (Supplementary Material) show different participants with different skin types and overall finger thickness at different positions of the finger successfully being imaged and processed to extract vasculature. 

In skin types I–IV, a certain degree of lateral scattering was seen with brightness being distributed away from the point where the light from the laser diode module is incident on the dorsal side of the finger. In previous studies, the images of the finger were taken from a device set up with LEDs and a near-infrared camera system but restricts the movement of the body site (the finger in most biometric applications) and is not a non-contact system \cite{Miura2004,Miura2005,Mulyono2008,Yang2014,Pham2015,Matsuda2017} except for visualisation systems \cite{Cuper2013}. Using a laser as an illumination source allows conservation of directionality of the light and using a CMOS camera in the visible-near-infrared region keeps the overall cost of the system low. 

\begin{figure*} [htp]
\begin{center}
\includegraphics[width=1.8\columnwidth]{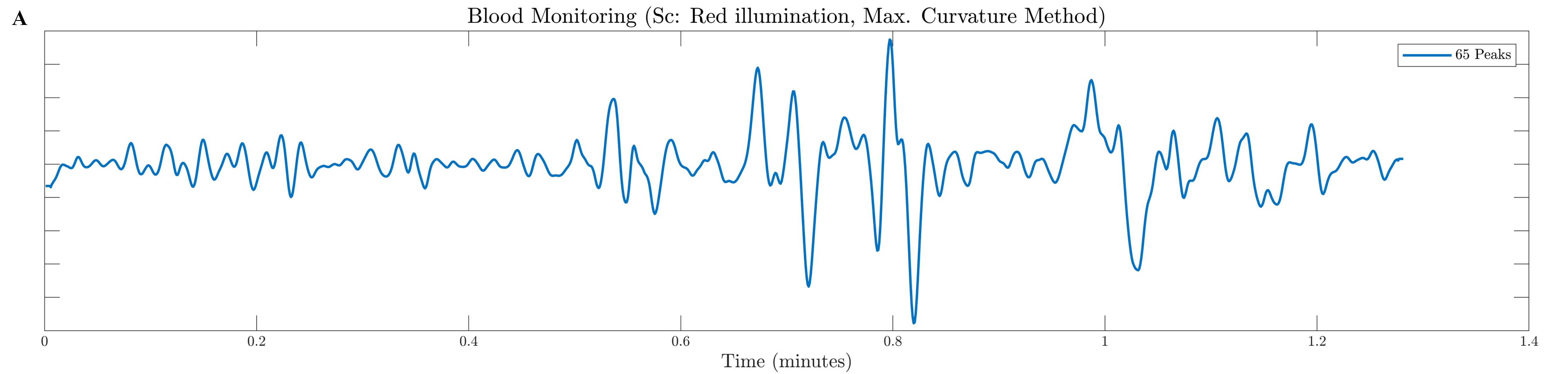} \\
\includegraphics[width=1.8\columnwidth]{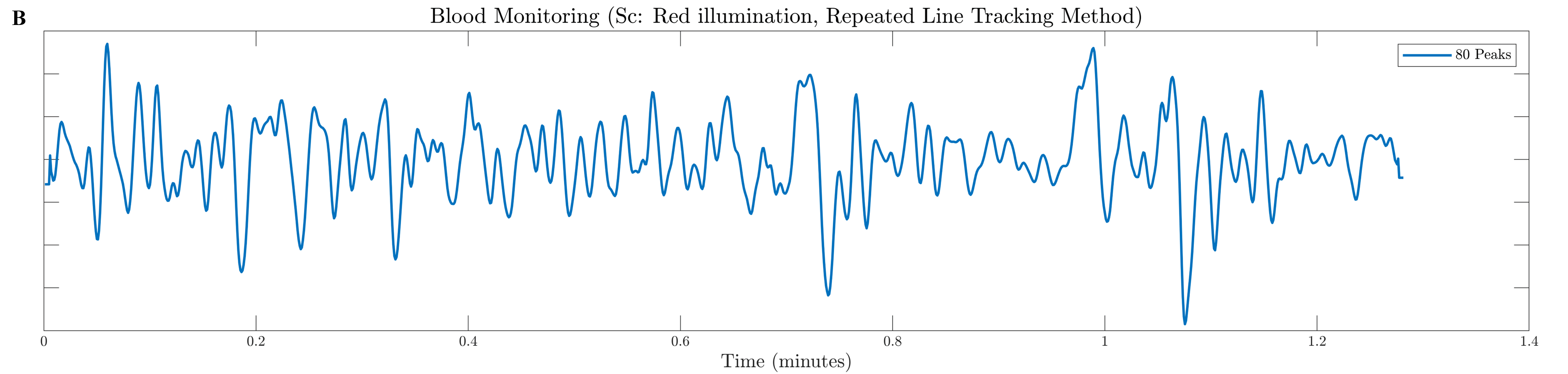} \\
\includegraphics[width=1.8\columnwidth]{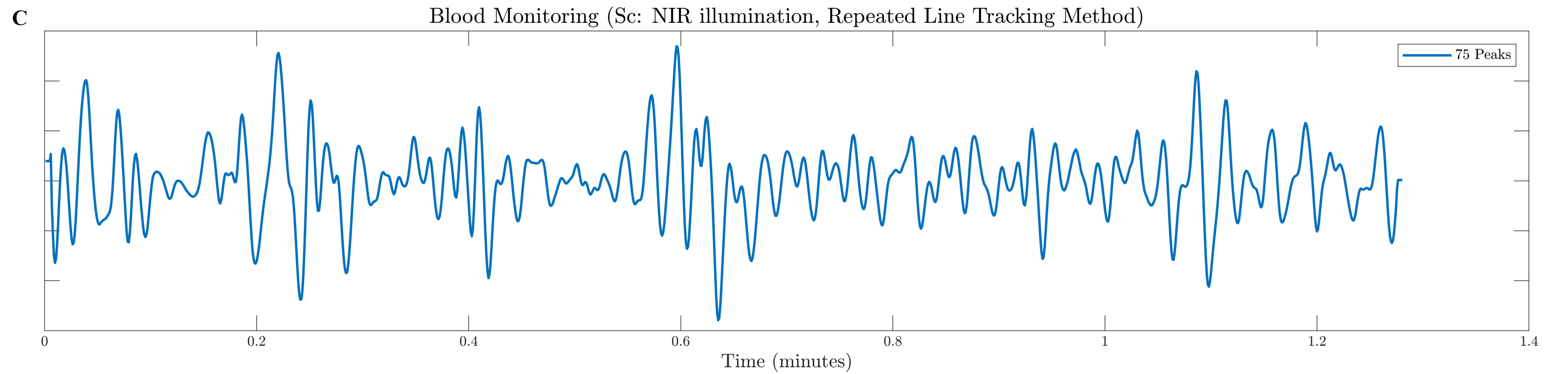} 
\end{center}
\caption{The results of vasculature mapping and monitoring for USP002 are compared for red illumination using both algorithms (A, B). However, the red illumination was not useful for all the participants. (C) shows the near-infrared illumination scenario combined with repeated line tracking
method to present a sinusoidal-like pattern. The overall detection and monitoring has been seen to be better with repeated line tracking as the pattern detection algorithm (C).}
\label{fig:Results_2}
\end{figure*}

We compare the results across methods, illumination wavelength and two of the participants (additional comparisons and the data is available upon reasonable request). In Figure \ref{fig:Results_1}, the final cardiac activity output compares both pattern mapping algorithms for the same video sequence captured from USP001. Both methods identify the vasculature and successfully monitor it to quantify the heart rate-related variations (with an margin of $+/-$ 5 bpm). For participant USP002, the comparison is made for red illumination with the maximum curvature method (A) and repeated line tracking method (B) in Figure \ref{fig:Results_2}. The use of light at red wavelengths with the capability of detecting vasculature could allow direct application of pulse oximetry principles to calculate the oxygen saturation in the blood in addition to the heart rate measurement, as seen in this case. However, the number of peaks (corresponding to beats per minute) are not identified with the same degree of accuracy when comparing the red illumination-maximum curvature method combination against other methods. The pulse oximeter measured a 77 beats for the imaging duration, which is within the margin of error of the other two instances, i.e. Figure \ref{fig:Results_2} (B, C). Red wavelength did not facilitate the identification and monitoring of blood vessel patterns with all the participants’ data but did work for USP002. For these results, the methodology was used to process and output the detected cardiac activity. Visually, the repeated line tracking method functions better in this scenario as well, with the identification of the blood vessel being a concern for the maximum curvature method. The pattern in Figure \ref{fig:Results_2} (B) can be fitted to a sinusoidal pattern better and the method successfully identifies the vein position in every frame of the data capture. 

In comparison, using near-infrared illumination source, the repeated line tracking algorithm performs much better and more consistently (Figure \ref{fig:Results_2} (C)) in comparison to the scenario using a red laser source. Using near-infrared illumination and repeated line tracking method for vasculature mapping prior to image processing performs better once again, and in other participants as well. This data, when fitted to a consistent pattern would replicate the familiar oximeter outputs seen by medical personnel from standard devices. Also, the number of peaks ($=75$) in the Figure \ref{fig:Results_2} (C) scenario corresponds to the heart rate of the participant, which was confirmed in this case with a pulse oximeter in the laboratory. Similarly, the previous scenarios were also compared and a $+/-$ 5 beats bpm margin was observed. Note that the acquisition scenarios of fig. \ref{fig:Results_2} (A, B) and fig. \ref{fig:Results_2} (C) are different with different illumination wavelengths and exposure times. An innate difference with the motion of the finger is also probable as identical positioning and movement are nearly impossible.

\section{Conclusion}
\label{sec:Conclusions}
We illustrated a simple and low-cost methodology for using a laser source and a CMOS-based imaging system for detecting and monitoring vascular activity. The algorithms for recognition of subsurface vasculature were adopted from the domain of biometric recognition and extended to monitor vascular activity. The methodology is not affected by motion of the finger. A successful interdisciplinary method was developed and tested on multiple, healthy participants. In this study, biometric pattern detection algorithms are successfully applied to diagnostics, enhancing the possibility for future research using low-cost equipment and non-contact methods. Specifically, the maximum curvature method performs better when identifying the vasculature pattern within the finger but its output does not provide a stable source for blood flow monitoring.

Repeated line tracking method, on the other hand, has a generally higher accuracy for eventual identification and monitoring vascular activity. The method’s outputs are far more robust to exposure time and finger movement variations. However, it is susceptible to noise which can be visibly seen in the outputs and requires consideration in post-processing. The ideal scenario for vascular activity monitoring within the scope of this study was found to be the higher exposure time, near-infrared illumination and repeated line tracking method. The overall system and method is affected by anatomical variations of skin type and finger thickness. In combination, the methodology is biased towards skin types I-IV in this current configuration using a visible camera with low sensitivity in the near-infrared regions. However, imaging with a near-infrared enhanced camera and cut-off filters has shown promise. While clinical application is the eventual goal, the current research illustrates the potential of a simple and low-cost approach to for point-of-care diagnostics. 

\section*{Acknowledgements}
We thank Dr Mark Finnis (Aeromechanical Systems Group, Cranfield University) for his support, providing the facilities for the laser laboratory for the experimental phase of the research. A special note of gratitude to all the participants for their patience. 

\section*{Author contributions}
The idea for this research was conceptualised by A.K., M.A.R. and D.B.J. The experiments and data analysis was performed by A.K. with feedback from D.B.J. All authors contributed towards writing the article and have read this article prior to submission to the journal. 

\section*{Financial disclosure}

None reported.

\section*{Conflict of interest}

The authors declare no potential conflict of interests.

\section*{Data Availability Statement}
The data that support the findings of this study are available from the corresponding author upon reasonable request.

\bibliography{Thesis_Ref}

\subsection*{Graphical Abstract Text}
Validated by both ease of measurement and the ongoing COVID-19 pandemic, low-cost and remote solutions of vital signs is essential for any diagnostic process. Inspired by novel biometric solutions, we illustrate a robust methodology that extracts vasculature distributions in the finger without motion artefacts. Further, we extend the approach to monitoring heart rate. Future work could provide valuable information in many fields, with postnatal monitoring being one such application. 

\subsection*{Graphical Abstract Figure}

\begin{figure}[h]
\begin{center}
\includegraphics[height=5cm]{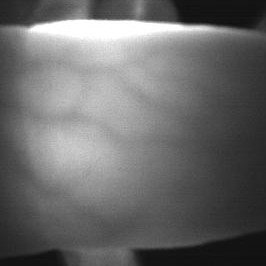}
\end{center}
\end{figure}

\end{document}